\documentclass[conference]{IEEEtran}

\usepackage[pdftex]{graphicx}
\usepackage{cite, balance}
\usepackage{epstopdf}
\usepackage{amssymb}
\usepackage[cmex10]{amsmath}
\usepackage{multirow}
\usepackage{algorithmic}
\usepackage{algorithm}
\usepackage[utf8]{inputenc}
\usepackage{soul, color}
\usepackage[caption=false,font=footnotesize]{subfig}

\begin{document}


\title{Diversity Combining for RF Energy Harvesting}

\author{
\IEEEauthorblockN{Dogay Altinel\IEEEauthorrefmark{1}\IEEEauthorrefmark{4}, Gunes Karabulut Kurt\IEEEauthorrefmark{1}}	
\IEEEauthorblockA{\IEEEauthorrefmark{1}Department of Electronics and Communication Engineering, Istanbul Technical University, Turkey}   
\IEEEauthorblockA{\IEEEauthorrefmark{4}Department of Electrical and Electronics Engineering, Istanbul Medeniyet University, Turkey
\\\ dogay.altinel@medeniyet.edu.tr, \{altineld, gkurt\}@itu.edu.tr}	
}


\maketitle

\begin{abstract} 
 
RF energy harvesting (RFEH) is a promising technology for energy requirements of wireless communication nodes. However, providing sufficient amount of energy to ensure self-sufficient devices based on RFEH may be challenging. In this paper, the use of diversity combining in RFEH systems is proposed to increase the amount of harvested energy. The power consumption of diversity combining process is also taken into account to analyze the net benefit of diversity combining. Performances of RFEH systems are investigated for selection combining (SC), equal gain combining (EGC), and maximal ratio combining (MRC) techniques. Simulations are conducted to compare the numerical results of SC, EGC, and MRC, and the results show that although the diversity combining techniques can improve the energy harvesting performance, the power consumption parameters have a critical importance while determining the suitable technique. 

\end{abstract}



\vspace{0.2cm}
\section{Introduction}
\label{sec:introduction}
 
Energy is a significant constraint for wireless communication devices that are generally dependent on limited capacity batteries. In recent years, energy harvesting has been investigated as an alternative energy source for low-power communication nodes. Energy harvesting circuits gather ambient energy from sunlight, vibration, air flow, thermal gradient or other types of harvestable energy sources. Similarly, energy harvesting from radio frequency (RF) signals is also considered as an alternative solution when utilized jointly with advanced semiconductor technology \cite{6951347}. In RF energy harvesting (RFEH) systems, a wireless node equipped with energy harvesting circuit captures radio signal from ambient by its antenna, then converts radio signal energy into direct current (DC) energy. The amount of harvested RF energy may not be sufficient to directly power-up the node. Therefore, the harvested energy from received DC signals is commonly used to recharge the battery of wireless node, which is utilized to  receive, process, and transmit information when required. In order to ensure sustainable communications, it is necessary to harvest the sufficient amount of energy and extend the lifetime of battery as required by the used application. In the literature, there are various studies investigating different ways to improve the amount of harvested RF energy. These approaches can be stated as special designs of circuit,  antenna, and signal types for RFEH systems as well as the use of multiple antennas, multiple frequency bands, and power management systems. In \cite{6159063}, focusing on circuit design, a new RFEH circuit is proposed to improve operational efficiency in the low incident power range. In \cite{5434266}, a compact aperture coupled patch rectenna is designed to receive arbitrarily polarized signals with high conversion efficiency. In \cite{5972612}, the use of multisine signal excitation is presented to increase the obtained DC power in energy harvesting systems. In \cite{6732976}, several distributed arrays of antennas, which are designed by scaling in array size, power, DC load, frequency, and gain level, are presented to increase harvested power and efficiency.  In \cite{6697364}, a triple-band antenna is developed to effectively harvest RF energy from available Wi-Fi  and cellular network frequency bands. In \cite{5395599}, microcontroller-based power management system is proposed to obtain more energy than direct connection.  

Although proposed approaches introduce RFEH systems with improved performances, they fail to make use of diversity combining techniques. In current wireless communication systems, diversity combining is considered as an important tool to enhance wireless link performance by alleviating the effects of radio channels' fading process \cite{4065786}. Diversity combining techniques are based on redundancy of transmitted information. Destinations combine multiple copies of the same information signal received over different points of a domain such as time, frequency, and space. While time and frequency diversity can be achieved in single-antenna transmission systems, space diversity needs the use of multiple antennas that are spaced sufficiently far apart  to obtain a diversity gain \cite{5039585}. On the receiver side, the used diversity combining technique plays an important role on the system performance. Three common diversity combining techniques are the selection combining (SC), the equal gain combining (EGC), and the maximal ratio combining (MRC). There is a trade-off between the performance and the complexity based on the selected diversity combining technique. These techniques are mainly used to increase signal-to-noise ratio (SNR) of the received signal. 
However, the aim in RFEH systems is to increase the amount of harvested RF energy instead of the value of received SNR. 

In order to benefit from diversity combining in RFEH systems, the design of an RFEH system employing a diversity combining technique is considered in this paper. In the literature, multiple-input multiple-output (MIMO) systems and energy beamforming at the transmitter side are studied to overcome propagation effects over distance in RFEH based new technologies. In \cite{6489506}, MIMO is used for maximizing the efficiency of simultaneous wireless information and power transfer (SWIPT) systems. In \cite{6568923,6954434,7274644}, wireless energy transfer, via energy beamforming from a multi-antenna transmitter to single-antenna receivers, is utilized in SWIPT systems, wireless powered communication (WPC) networks, and  backscatter communication systems, respectively. These works give useful insights on the improvement of RFEH technology, but they do not handle the use of diversity combining techniques at the receiver side. 

Our motivation in this paper is to increase the amount of energy obtained from RF signals by using diversity combining, and to analyze the performance of diversity combining based RFEH systems according to the selected technique. We study the problem of diversity combining in a communication network, where the receiver is equipped with multiple antennas. We propose a new receiver structure making use of diversity combining to improve the performance of RFEH systems. We also take the power consumption of diversity combining process into account. We analyze three diversity combining techniques, which are SC, EGC, and MRC. We compare the performances of RFEH systems in terms of the harvested power and the net obtained power. It is shown via simulation results that diversity combining techniques can provide significant benefit for RFEH systems.


\section{System Model and Receiver Structure}

\begin{figure}[t]
\centering
\includegraphics[width=0.42\textwidth]{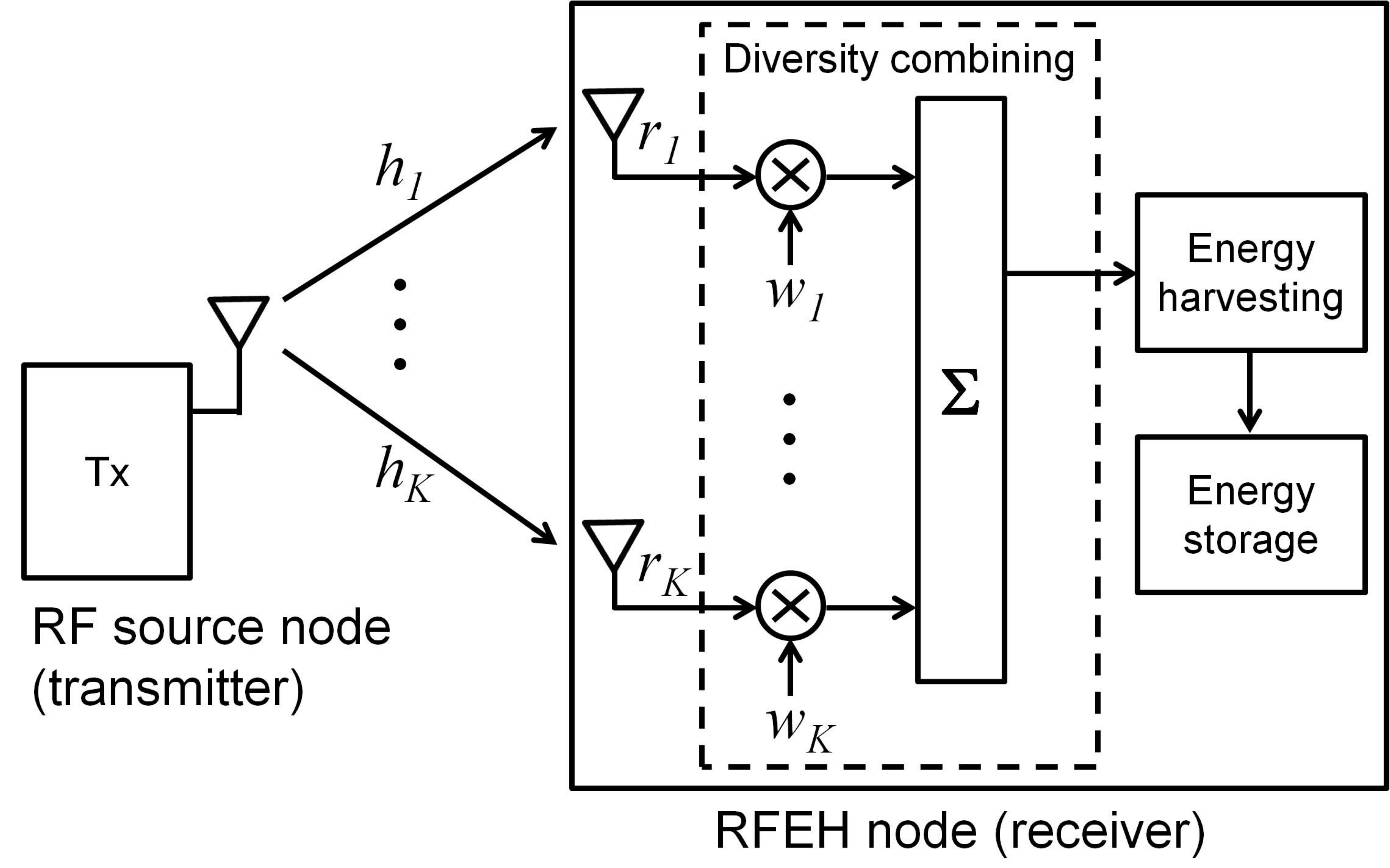}
\caption{System model including an RF source node (transmitter) and an RFEH node (receiver) with diversity combining.}
\label{fig:systemmodel1}
\end{figure}

We consider a system composed of an RF source-harvesting node (transmitter-receiver) pair with single transmit antenna and multiple receive antennas, respectively, as illustrated in Fig. \ref{fig:systemmodel1}. The RF source node sends energy in the transmitted symbol to the RFEH node through wireless channels. The RFEH node contains a diversity combining unit, energy harvesting unit, and energy storage unit after antennas. The received signals are combined in the diversity combining unit, 
and harvested in the energy harvesting unit. The harvested energy is stored in the energy storage unit. Assuming $K$ receive antennas and flat fading channels, the  received signal on the $k^ \textrm{th}$ antenna can be modeled as 
\begin{eqnarray}
r_k = \sqrt{P_{t}} h_k x + z_k,  \;k = 1, 2,\ldots, K,
\label{channelmodel}
\end{eqnarray}
where $x$ represents the transmitted symbol. $z_k$ is the complex additive white Gaussian noise (AWGN) with zero mean and $\sigma^2_z$ variance. $h_k$ is the channel coefficient affecting the transmitted symbol between the RF source node and the RFEH node,  $h_k=|h_k| e^{j \theta_k}$, where $|h_k|$ and $\theta_k$ are amplitude and phase of $h_k$, respectively. $P_{t}$ denotes the average transmit power. It is assumed that the transmitted symbol is normalized in terms of power, where its expected value becomes unity, $E[|x|^2] = 1$. Depending on the related channel coefficient, the received power from $k^\textrm{th}$ antenna $(P_{r,k})$ becomes
\begin{eqnarray}
P_{r,k}= P_{t} |h_k|^2. 
\label{receivedpower1}
\end{eqnarray}
The harvested power from $k^\textrm{th}$ antenna $(P_{h,k})$ can be expressed as \cite{6623062}
\begin{eqnarray}
P_{h,k}=\eta P_{t} |h_k|^2,
\label{harvestedpower1}
\end{eqnarray}
where $\eta$, $0<\eta<1$, is the conversion efficiency of RF signal to DC signal corresponding to the performance of the energy harvesting unit in the RFEH node. 
The harvested energy from $k^\textrm{th}$ antenna $(E_{h,k})$ throughout the harvesting time $(T)$ is stated as
\begin{eqnarray}
E_{h,k}=T \eta P_{t} |h_k|^2.  
\label{harvestedenergy1}
\end{eqnarray}

In the proposed receiver, we target to increase the amount of harvested RF power by using diversity combining techniques. In that case, the weighted form of received signal can be modeled as
\begin{eqnarray}
y_k = \sqrt{P_{t}} w_k h_k x + w_k z_k,  \quad k = 1, 2,\ldots, K,
\label{channelmodel}
\end{eqnarray}
where $w_k$ is the combiner weight coefficient corresponding to the antenna branch $k$, $w_k \in \mathbb{C}$. For the rest of this paper, we target the power expressions, which can be stated as
\begin{eqnarray}
P_{h,k}=\eta P_{t} |w_k h_k|^2,
\label{harvestedpower3}
\end{eqnarray}
denoting the harvested power from $k^\textrm{th}$ antenna and
\begin{eqnarray}
P_{h}= \eta P_{t} \left|\sum\limits_{k=1}^K w_k h_k\right|^2, 
\label{harvestedpower4}
\end{eqnarray}
denoting the total harvested power in the presence of diversity combiner. 

Considering linear combination of incoming powers, the total maximum harvestable power $(P_{h,{max}})$ 
can be obtained as
\begin{eqnarray}
P_{h,{max}}= \sum\limits_{k=1}^K \eta P_{t} |h_k|^2. 
\label{harvestedpower2}
\end{eqnarray}
In the next section, we examine diversity combining techniques and their impacts on the harvested power.

\section{Analysis of Diversity Combining for RFEH}

Diversity is an effective way for increasing the received SNR in the presence of wireless fading channels. Space diversity, having additional antennas,  increases the probability that the value of SNR is sufficiently high at the receiver. Here, based on the system model including diversity combining, we focus on the harvested RF power instead of the received SNR. Unlike the use of diversity combining to maximize SNR, the use of diversity combining in RFEH targets to increase the net obtained power. Therefore, we need to take the power consumption of diversity combining process into account and analyze the net benefit of diversity combining. For any diversity combining technique, the power consumption of $k^\textrm{th}$ antenna branch $(P_{c,k})$ can be modeled as 
\begin{eqnarray}
P_{c,k}= \beta P_{w,k} + P_{f,k}, 
\label{powerconsumption11}
\end{eqnarray}
where $P_{w,k}$ is the power of combiner weight coefficient at $k^\textrm{th}$ antenna branch. It is dependent on the received signal. The parameter $\beta$ denotes the combiner weight efficiency, as a real number $\beta \ge 1$, which encompasses the impact of power losses in the generation of combiner weight coefficient. The value of $\beta$ changes according to the type of diversity combining technique. $P_{f,k}$ represents the constant power consumption due to the circuit elements at $k^\textrm{th}$ antenna branch, which is not dependent on the received signal.

Considering the overall diversity combining system, the total power consumption $(P_c)$ can be formulated as
\begin{eqnarray}
P_c= \beta P_w + P_d, 
\label{powerconsumption1a}
\end{eqnarray}
where $P_w$ is the total power of combiner weight coefficients. It can be calculated as 
\begin{eqnarray}
P_w= \sum\limits_{k=1}^K \: P_{w,k} = \sum\limits_{k=1}^K \: |w_k|^2. 
\label{pw1}
\end{eqnarray}
$P_d$ in (\ref{powerconsumption1a})  denotes the total constant power consumption of diversity combining system. The value of $P_d$ depends on the type of receiver beamforming scheme that can be classified as analog, digital, and hybrid beamforming \cite{abbas2016towards}. In general terms, $P_d$ includes $P_{f,k}$ values in all antenna branches and the power consumption of summation unit $(P_{s})$. It can be expressed as 
\begin{eqnarray}
P_d= \sum\limits_{k=1}^K \: P_{f,k} + P_{s}. 
\label{pd1}
\end{eqnarray}
It is reasonable to assume that each antenna branch has the same circuit design. In that case, $P_{f,k}$ values become equal, $P_{f,k}=P_f$, $\forall k$. The definition of $P_d$ becomes   
\begin{eqnarray}
P_d= K P_{f} + P_{s}. 
\label{pd2}
\end{eqnarray}
Considering both the total harvested power in (\ref{harvestedpower4}) and the total power consumption, the net obtained power $(P_{net})$ for all type of receiver beamforming schemes can be expressed  as       
\begin{eqnarray}
P_{net}= \eta P_{t} \left|\sum\limits_{k=1}^K w_k h_k\right|^2 - \beta \sum\limits_{k=1}^K \: |w_k|^2 - P_d. 
\label{netobtainedpower1}
\end{eqnarray}
Note that $\eta$, $\beta$, and $P_d$ depend on the receiver circuits, and $|h_k|$ depends on the operation environment. These parameters can be considered as constant for a given system. To improve the net obtained power, the value of $P_t$ can be increased. The value of $P_t$ is set on the transmitter side. However, in the receiver side, the combiner weight coefficients can be optimized. According to equation (\ref{netobtainedpower1}), the optimization problem (\textbf{P1}) can be expressed as 
\begin{eqnarray}
(\textbf{P1}): \quad
\begin{aligned}
& \underset{\textbf{w}}{\text{max}}
& & \eta P_{t} \left|\sum\limits_{k=1}^K w_k h_k\right|^2 - \beta \sum\limits_{k=1}^K \: |w_k|^2 - P_d,
\end{aligned}
\label{opt1}
\end{eqnarray}
where $\textbf{w}$ is the vector of combiner weight coefficients as $\textbf{w} =[w_1,w_2,\ldots,w_K]$. As the solution of decision variables in the nonlinear multivariable optimization problem (\textbf{P1}), the $w_k$ values in the vector $\textbf{w}$ go to infinity. Because, the  value of $P_{net}$ increases as the values of $w_k$ increase. Actually it is not realistic to use such a solution for actual receivers. Therefore, it is necessary to limit the total power of combiner weight coefficients, i.e. the value of $P_w$. In that case, the optimization problem (\textbf{P2}) is stated as
\begin{eqnarray}
(\textbf{P2}): \quad
\begin{aligned}
& \underset{\textbf{w}}{\text{max}}
& & \eta P_{t} \left|\sum\limits_{k=1}^K w_k h_k\right|^2 - \beta \sum\limits_{k=1}^K \: |w_k|^2 - P_d \\ 
& \text{s.t.}
& & \sum\limits_{k=1}^K \: |w_k|^2  \leq \xi,
\end{aligned}
\label{opt2}
\end{eqnarray}
where $\xi$ is the limit value of $P_w$ as a real-valued constant ($\xi \in \mathbb{R}$). Since we focus on diversity combining for the harvested power and energy unlike diversity combining for the received SNR, it is necessary to take the conservation of energy into account. The value of output power can not exceed the value of input power. It requires the expression of 
\begin{eqnarray}
\sum\limits_{k=1}^K \: |w_k|^2  \le 1 
\label{conservation1}
\end{eqnarray}
for actual networks, i.e. $\xi=1$, \cite{schwartz}. The solution of constrained nonlinear multivariable optimization problem (\textbf{P2}) can be obtained in different forms based on the knowledge of amplitude and phase of channel coefficients. In this context, we analyze MRC, EGC, and SC seperately.

\subsection{Maximal Ratio Combining}

In the receiver with MRC, all received signals are weighted and combined coherently to maximize the value of SNR. The output gives the sum of all individual SNRs through the antennas. Here, both amplitude and phase values of channel coefficient at each branch need to be correctly estimated for all instances in time. In our case, the same process is conducted to maximize the net obtained power. The solution of optimization problem (\textbf{P2}) for any value of $\xi$ can be found by using an optimization algorithm such as the interior-point algorithm. In this solution, all received signals are combined that the phases of signals take the same fixed value (not necessary to be zero) to obtain the maximum power. Without loss of generality, the phase of signals can be fixed to zero. In that case, according to the Schwarz inequality, we observe that
\begin{eqnarray}
\left| \sum\limits_{k=1}^K \: w_k h_k \right|^2 \le \sum\limits_{k=1}^K \: |w_k|^2 \sum\limits_{k=1}^K \: |h_k|^2, 
\label{schwartz1}
\end{eqnarray}
hence the solution for $w_k$ becomes
\begin{eqnarray}
w_k= c h^*_k, \quad \forall k, 
\label{wgeneral}
\end{eqnarray}
where $c$ is a real-valued constant, $c \in \mathbb{R}$, depending on the value of $\xi$. For $\xi=1$, the combiner weight coefficients can be expressed as
\begin{eqnarray}
w_k= \frac{h^*_k}{||\textbf{h}||} = \frac{|h_k|}{||\textbf{h}||} e^{-j \theta_k}. 
\label{wmrc}
\end{eqnarray}
where $||\textbf{h}||= \sqrt{ \sum\limits_{k=1}^K \: |h_k|^2 }$, i.e. $c=1/||\textbf{h}||$. The normalization of $P_w$ ensures both the conservation of energy and the solution of (\textbf{P2}). Considering $P_w=1$,  the harvested power for MRC $(P_{h,mrc})$ is calculated as 
\begin{eqnarray}
P_{h,mrc}= \eta P_{t} \sum\limits_{k=1}^K |h_k|^2, 
\label{netobtainedpower2e}
\end{eqnarray}
and the net obtained power for MRC $(P_{net,mrc})$ becomes
\begin{eqnarray}
P_{net,mrc}= \eta P_{t} \sum\limits_{k=1}^K |h_k|^2 - \beta_{mrc} - P_{d,mrc}, 
\label{netobtainedpower2}
\end{eqnarray}
where $\beta_{mrc}$ and $P_{d,mrc}$ are the related values of $\beta$ and $P_d$ for MRC, respectively. Note that equation (\ref{netobtainedpower2}) includes the total maximum harvestable power expression in (\ref{harvestedpower2}). 

\subsection{Equal Gain Combining}

The receiver using EGC combines the received signals coming from all antennas coherently to increase the value of received SNR. In our case, the same process is conducted for improving the net obtained power. The combiner weight coefficients can be written as
\begin{eqnarray}
w_k= |w_k| e^{-j \arg\{h_k\}} = \gamma e^{-j \theta_k}, 
\label{wegc}
\end{eqnarray}
where $\gamma $ denotes the amplitude of $w_k$, which is constant for all $k$ values. Here, only the knowledge of phase values of channel coefficients is sufficient, amplitude values of channel coefficients are not required.  

In order to ensure conservation of energy and equal gain approach for each branch, $\gamma$ is set to $\gamma=1/ \sqrt{K}$. In that case, the harvested power for EGC $(P_{h,egc})$ is expressed as
\begin{eqnarray}
P_{h,egc}= \frac{\eta P_{t}}{K} \left( \sum\limits_{k=1}^K \: |h_k| \right)^2, 
\label{harvestedenergy_egc_e}
\end{eqnarray}
and the net obtained power for EGC $(P_{net,egc})$ becomes
\begin{eqnarray}
P_{net,egc}= \frac{\eta P_{t}}{K} \left( \sum\limits_{k=1}^K \: |h_k| \right)^2 - \beta_{egc} - P_{d,egc}, 
\label{harvestedenergy_egc}
\end{eqnarray}
where $\beta_{egc}$ and $P_{d,egc}$ are the related values of $\beta$ and $P_d$ for EGC, respectively. $\beta_{egc}$ is expected to be smaller than $\beta_{mrc}$ due to relative implementation simplicity of EGC.

\subsection{Selection Combining}

In SC, the simplest combining technique, only a signal received by a single antenna is used for an instance in time. The receiver employing SC measures the SNRs of all signals coming from the antennas and uses the signal with the maximum SNR. In our case, the receiver selects the antenna with the maximum received power. The combiner weight coefficients are assigned as 
\begin{eqnarray}
w_k=
\begin{cases} 
1,& k = \: $arg$ \: \underset{k}{\text{max}} \: (P_{r,k}) \\ 0,& $otherwise$.
\end{cases}
\label{wsc}
\end{eqnarray}
(\ref{wsc}) implies that only the knowledge of amplitude values of channel coefficients is sufficient, phase values of channel coefficients are not required. The harvested power for SC $(P_{h,sc})$ is calculated as
\begin{eqnarray}
P_{h,sc}= \underset{k}{\text{max}} \: (\eta P_{t} |h_k|^2). 
\label{harvestedenergy_sc_e}
\end{eqnarray}
Note that $\beta$ value for SC is zero, $\beta_{sc}=0$, since there are no generated signals for the combiner weight coefficients in the real circuits. The net obtained power in case of SC $(P_{net,sc})$ can be stated as
\begin{eqnarray}
P_{net,sc}= \underset{k}{\text{max}} \: (\eta P_{t} |h_k|^2) - P_{d,sc}, 
\label{harvestedenergy_sc}
\end{eqnarray}
where $P_{d,sc}$ is the value of $P_d$ for SC. Unlike $P_{d,mrc}$ and $P_{d,egc}$, $P_{d,sc}$ contains only the constant power consumption due to the circuit elements at $k^\textrm{th}$ antenna branch. It can be expressed as
\begin{eqnarray}
P_{d,sc}= P_{f,k}, \quad k= \text{arg} \: \underset{k}{\text{max}} \: (T \eta P_{t} |h_k|^2),
\label{pd_sc}
\end{eqnarray}
where $k \in \{1,2,\ldots,K\}$.

\begin{figure*}[t]
\centering
\begin{minipage}{.45\linewidth}
    \includegraphics[width=\linewidth]{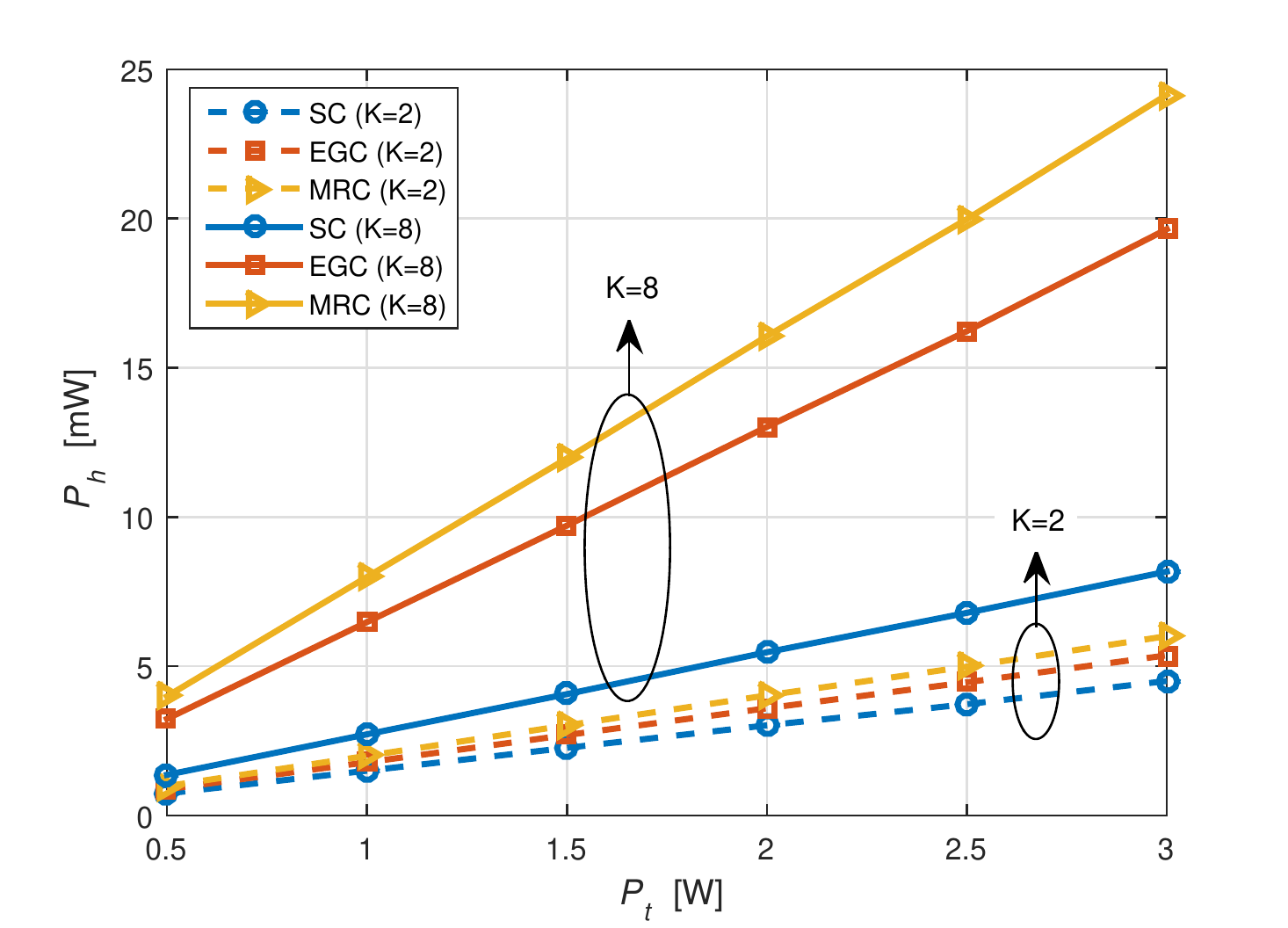}
    \caption{Mean values of the harvested power for SC, EGC, and MRC are plotted vs. the average transmit power, where the values are illustrated by dashed line for $K=2$ and solid line for $K=8$.}
    \label{sim1a}
\end{minipage}
\hfill
\begin{minipage}{.45\linewidth}
    \includegraphics[width=\linewidth]{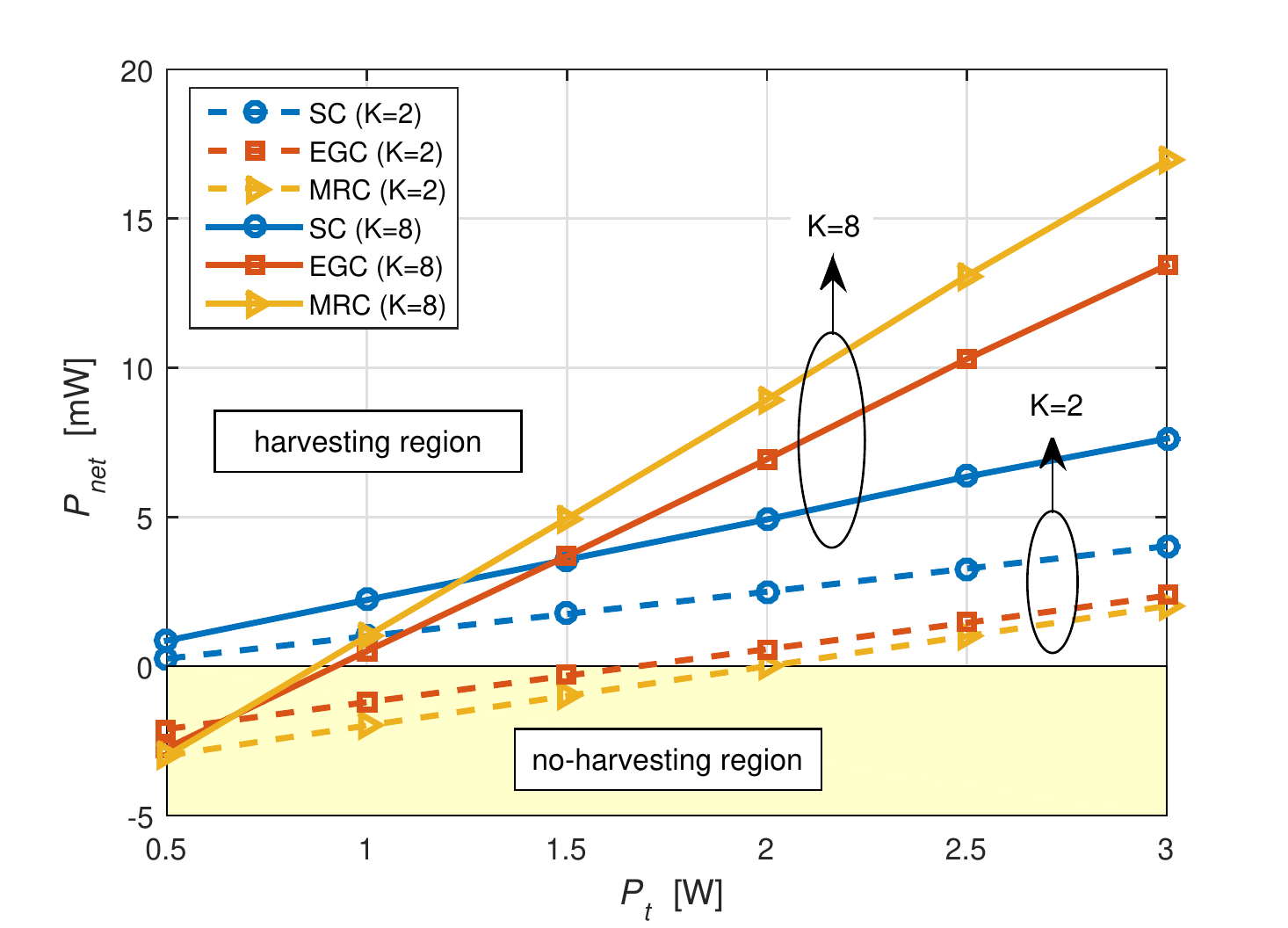}
    \caption{Mean values of the net obtained power for SC, EGC, and MRC are plotted vs. the average transmit power. No-harvesting region is illustrated by yellow color, where energy harvesting is not achieved.}
    \label{sim2a}
\end{minipage}
\end{figure*}

\section{Numerical Results}

Numerical results of simulations for different diversity combining techniques are presented in a comparative fashion in order to show the performances of diversity combining techniques. The channel coefficients are assumed to be Rayleigh distributed with a variance that is equal to the path loss. 
Without loss of generality, the path loss is considered as $10^{-3}$. The mean values of harvested power for SC, EGC, and MRC are illustrated for $K=2$ and $K=8$ in Fig. \ref{sim1a}.  It is seen that the increase of the number of receive antennas improves the amount of harvested power significantly for all diversity combining techniques. MRC is the best diversity combining technique for the harvested power. Note that although SC shows the worst performance for each fixed $K$ value, SC for $K=8$ provides better performance than MRC for $K=2$.

The power consumption of each unit changes based on the design of receiver circuits. It is possible to design low-power integrated diversity receivers as studied in \cite{735549,5477304}. As MRC is more complex than EGC, the combiner weight efficiency parameters are selected as $\beta_{mrc}=2$ and $\beta_{egc}=1$. Other parameters are set to $P_{s}=1$ mW and $P_{f,k}=0.5$ mW $\forall k$ to evaluate the performances depending on the net obtained power. In Fig. \ref{sim2a}, the mean values of the net obtained power vs. the average transmit power are demonstrated. There are two regions for the negative and positive values of the net obtained power in the figure, named as no-harvesting region and harvesting region, respectively. In the no-harvesting region, energy harvesting is not achieved. In the harvesting region, energy harvesting can be achieved with an appropriate diversity combining technique that provides the best performance. In case of $K=2$, the results show that energy harvesting for EGC and MRC is not possible in the ranges of [$0,1.6$] and [$0,1.9$] W, respectively. It is also seen that SC is the best technique for $K=2$. However, in case of $K=8$, SC outperforms other techniques up to $P_t=1.2$ mW. MRC is the best technique for higher values of $P_t$. Note that the performance of MRC for the net obtained power is not the best for all regions, although the performance of MRC for the harvested power becomes the best. 

%
%
%

\vspace{0.2cm}
\section{Conclusion}
\label{sec:conclusion}
The use of diversity combining is proposed to improve the performance of RFEH systems. The received RF signals are weighted and combined by the diversity combiner in the receiver. The output of diversity combiner is introduced as the input of energy harvester. The harvested energy is stored in an energy storage unit. Based on the proposed receiver structure, RFEH performances are analyzed for SC, EGC, and MRC techniques. The net obtained power depends on the power consumptions of circuits during combining process and changes with the type of diversity combining technique. Numerical results show that diversity combining provides significant increase in the amount of harvested energy. Moreover, the selection of diversity combining technique based on the operating region plays an important role on the performance of RFEH systems.


\balance

\end{document}